\documentclass[12pt]{article}
\usepackage[ansinew]{inputenc} 
\usepackage[english]{babel} 
\begin{document}
\def\hybrid{\topmargin 0pt      \oddsidemargin 0pt
        \headheight 0pt \headsep 0pt
        \textheight 9in         
        \textwidth 6.25in       
        \marginparwidth .875in
        \parskip 5pt plus 1pt   \jot = 1.5ex}
\catcode`\@=11
\def\marginnote#1{}
\hyphenation{bo-so-ni-zed}
\newcount\hour
\newcount\minute
\newtoks\amorpm
\def\draftlabel#1{{\@bsphack\if@filesw {\let\thepage\relax
   \xdef\@gtempa{\write\@auxout{\string
      \newlabel{#1}{{\@currentlabel}{\thepage}}}}}\@gtempa
   \if@nobreak \ifvmode\nobreak\fi\fi\fi\@esphack}
        \gdef\@eqnlabel{#1}}
\def\@eqnlabel{}
\def\@vacuum{}
\def\draftmarginnote#1{\marginpar{\raggedright\scriptsize\tt#1}}
\def\draft{\oddsidemargin -.5truein
        \def\@oddfoot{\sl preliminary draft \hfil
        \rm\thepage\hfil\sl\today\quad\militarytime}
        \let\@evenfoot\@oddfoot \overfullrule 3pt
        \let\label=\draftlabel
        \let\marginnote=\draftmarginnote

\def\@eqnnum{(\theequation)\rlap{\kern\marginparsep\tt\@eqnlabel}%
\global\let\@eqnlabel\@vacuum}  }


\def\numberbysection{\@addtoreset{equation}{section}
        \def\theequation{\thesection.\arabic{equation}}}

\def\underline#1{\relax\ifmmode\@@underline#1\else
 $\@@underline{\hbox{#1}}$\relax\fi}

\catcode`@=12 \relax

\numberbysection

\topmargin 0pt \advance \topmargin by -\headheight \advance \topmargin by -\headsep

\textheight 8.9in

\oddsidemargin 0pt \evensidemargin \oddsidemargin \marginparwidth 0.5in

\textwidth 6.5in

\topmargin -.6in

\def\rh{{\hat \rho}}
\def\alie{{\hat{\cal G}}}
\def\sect#1{\section{#1}}

\def\sd{D\!\!\!/}
\def\sp{\partial\!\!\!/}
\def\sa{A\!\!\!/}
\def\sb{b\!\!\!/}
\def\ss{s\!\!\!/}
\def\rf#1{(\ref{#1})}
\def\lab#1{\label{#1}}
\def\nonu{\nonumber}
\def\br{\begin{eqnarray}}
\def\er{\end{eqnarray}}
\def\be{\begin{equation}}
\def\ee{\end{equation}}
\def\eq{\!\!\!\! &=& \!\!\!\! }
\def\foot#1{\footnotemark\footnotetext{#1}}
\def\lb{\lbrack}
\def\rb{\rbrack}
\def\llangle{\left\langle}
\def\rrangle{\right\rangle}
\def\blangle{\Bigl\langle}
\def\brangle{\Bigr\rangle}
\def\llbrack{\left\lbrack}
\def\rrbrack{\right\rbrack}
\def\lcurl{\left\{}
\def\rcurl{\right\}}
\def\({\left(}
\def\){\right)}
\newcommand{\nit}{\noindent}
\newcommand{\ct}[1]{\cite{#1}}
\newcommand{\bi}[1]{\bibitem{#1}}
\def\lskip{\vskip\baselineskip\vskip-\parskip\noindent}
\relax

\def\tr{\mathop{\rm tr}}
\def\Tr{\mathop{\rm Tr}}
\def\v{\vert}
\def\bv{\bigm\vert}
\def\Bgv{\;\Bigg\vert}
\def\bgv{\bigg\vert}
\newcommand\partder[2]{{{\partial {#1}}\over{\partial {#2}}}}
\newcommand\funcder[2]{{{\delta {#1}}\over{\delta {#2}}}}
\newcommand\Bil[2]{\Bigl\langle {#1} \Bigg\vert {#2} \Bigr\rangle}  
\newcommand\bil[2]{\left\langle {#1} \bigg\vert {#2} \right\rangle} 
\newcommand\me[2]{\left\langle {#1}\bv {#2} \right\rangle} 
\newcommand\sbr[2]{\left\lbrack\,{#1}\, ,\,{#2}\,\right\rbrack}
\newcommand\pbr[2]{\{\,{#1}\, ,\,{#2}\,\}}
\newcommand\pbbr[2]{\lcurl\,{#1}\, ,\,{#2}\,\rcurl}
%
\def\a{\alpha}
\def\at{{\tilde A}^R}
\def\atc{{\tilde {\cal A}}^R}
\def\atcm#1{{\tilde {\cal A}}^{(R,#1)}}
\def\b{\beta}
\def\btil{{\tilde b}}
\def\dc{{\cal D}}
\def\d{\delta}
\def\D{\Delta}
\def\eps{\epsilon}
\def\bareps{{\bar \epsilon}}
\def\vareps{\varepsilon}
\def\fptil{{\tilde F}^{+}}
\def\fmtil{{\tilde F}^{-}}
\def\gh{{\hat g}}
\def\g{\gamma}
\def\G{\Gamma}
\def\grad{\nabla}
\def\h{{1\over 2}}
\def\l{\lambda}
\def\L{\Lambda}
\def\m{\mu}
\def\n{\nu}
\def\o{\over}
\def\om{\omega}
\def\O{\Omega}
\def\p{\phi}
\def\P{\Phi}
\def\pa{\partial}
\def\pr{\prime}
\def\pt{{\tilde \Phi}}
\def\qs{Q_{\bf s}}
\def\ra{\rightarrow}
\def\s{\sigma}
\def\S{\Sigma}
\def\t{\tau}
\def\th{\theta}
\def\Th{\Theta}
\def\tpp{\Theta_{+}}
\def\tmm{\Theta_{-}}
\def\tpg{\Theta_{+}^{>}}
\def\tms{\Theta_{-}^{<}}
\def\tp0{\Theta_{+}^{(0)}}
\def\tm0{\Theta_{-}^{(0)}}
\def\ti{\tilde}
\def\wti{\widetilde}
\def\jc{J^C}
\def\bj{{\bar J}}
\def\sj{{\jmath}}
\def\bsj{{\bar \jmath}}
\def\bp{{\bar \p}}
\def\vp{\varphi}
\def\vt{{\tilde \varphi}}
\def\faa{Fa\'a di Bruno~}
\def\ca{{\cal A}}
\def\cb{{\cal B}}
\def\ce{{\cal E}}
\def\cg{{\cal G}}
\def\cgh{{\hat {\cal G}}}
\def\ch{{\cal H}}
\def\chh{{\hat {\cal H}}}
\def\cl{{\cal L}}
\def\cm{{\cal M}}
\def\cn{{\cal N}}
\def\ns{N_{{\bf s}}}
\newcommand\sumi[1]{\sum_{#1}^{\infty}}   
\newcommand\fourmat[4]{\left(\begin{array}{cc}  
{#1} & {#2} \\ {#3} & {#4} \end{array} \right)}

%
\def\lie{{\cal G}}
\def\kmlie{{\hat{\cal G}}}
\def\dlie{{\cal G}^{\ast}}
\def\elie{{\widetilde \lie}}
\def\edlie{{\elie}^{\ast}}
\def\hlie{{\cal H}}
\def\flie{{\cal F}}
\def\wlie{{\widetilde \lie}}
\def\f#1#2#3 {f^{#1#2}_{#3}}
\def\winf{{\sf w_\infty}}
\def\win1{{\sf w_{1+\infty}}}
\def\hwinf{{\sf {\hat w}_{\infty}}}
\def\Winf{{\sf W_\infty}}
\def\Win1{{\sf W_{1+\infty}}}
\def\hWinf{{\sf {\hat W}_{\infty}}}
\def\Rm#1#2{r(\vec{#1},\vec{#2})}          
\def\OR#1{{\cal O}(R_{#1})}           
\def\ORti{{\cal O}({\widetilde R})}           
\def\AdR#1{Ad_{R_{#1}}}              
\def\dAdR#1{Ad_{R_{#1}^{\ast}}}      
\def\adR#1{ad_{R_{#1}^{\ast}}}       
\def\KP{${\rm \, KP\,}$}                 
\def\KPl{${\rm \,KP}_{\ell}\,$}         
\def\KPo{${\rm \,KP}_{\ell = 0}\,$}         
\def\mKPa{${\rm \,KP}_{\ell = 1}\,$}    
\def\mKPb{${\rm \,KP}_{\ell = 2}\,$}    
%
\def\rlx{\relax\leavevmode}
\def\inbar{\vrule height1.5ex width.4pt depth0pt}
\def\IZ{\rlx\hbox{\sf Z\kern-.4em Z}}
\def\IR{\rlx\hbox{\rm I\kern-.18em R}}
\def\IC{\rlx\hbox{\,$\inbar\kern-.3em{\rm C}$}}
\def\IN{\rlx\hbox{\rm I\kern-.18em N}}
\def\IO{\rlx\hbox{\,$\inbar\kern-.3em{\rm O}$}}
\def\IP{\rlx\hbox{\rm I\kern-.18em P}}
\def\IQ{\rlx\hbox{\,$\inbar\kern-.3em{\rm Q}$}}
\def\IF{\rlx\hbox{\rm I\kern-.18em F}}
\def\IG{\rlx\hbox{\,$\inbar\kern-.3em{\rm G}$}}
\def\IH{\rlx\hbox{\rm I\kern-.18em H}}
\def\II{\rlx\hbox{\rm I\kern-.18em I}}
\def\IK{\rlx\hbox{\rm I\kern-.18em K}}
\def\IL{\rlx\hbox{\rm I\kern-.18em L}}
\def\one{\hbox{{1}\kern-.25em\hbox{l}}}
\def\0#1{\relax\ifmmode\mathaccent"7017{#1}%
B        \else\accent23#1\relax\fi}
\def\omz{\0 \omega}
%
\def\ltimes{\mathrel{\vrule height1ex}\joinrel\mathrel\times}
\def\rtimes{\mathrel\times\joinrel\mathrel{\vrule height1ex}}
%
\def\mark{\noindent{\bf Remark.}\quad}
\def\prop{\noindent{\bf Proposition.}\quad}
\def\theor{\noindent{\bf Theorem.}\quad}
\def\name{\noindent{\bf Definition.}\quad}
\def\exam{\noindent{\bf Example.}\quad}
\def\proof{\noindent{\bf Proof.}\quad}
                %
                %
\def\SCAND#1#2#3{{\sl Scandinavian Journal of Statistics} {\bf #1} (#2) #3}
\def\AS#1#2#3{{\sl Applied Statistics} {\bf #1} (#2) #3}
\def\METRO#1#2#3{{\sl Metrologia} {\bf #1} (#2) #3}
\def\CHEMO#1#2#3{{\sl Journal of Chemometrics} {\bf #1} (#2) #3}
\def\AQA#1#2#3{{\sl Accreditation and Quality Assurance} {\bf #1} (#2) #3}
\def\QN#1#2#3{{\sl Química Nova} {\bf #1} (#2) #3}
\def\BIO#1#2#3{{\sl Biometrika} {\bf #1} (#2) #3}
\def\TEC#1#2#3{{\sl Techno\-me\-trics} {\bf #1} (#2) #3}
\def\FRE#1#2#3{{\sl Fresenius J Anal Chem} {\bf #1} (#2) #3}
\def\ISR#1#2#3{{\sl International Statistical Review} {\bf #1} (#2) #3}
\def\JASA#1#2#3{{\sl Journal of the American Statistical Association} {\bf #1} (#2) #3}
                %
                \def\a{\alpha}
                \def\b{\beta}
                \def\ca{{\cal A}}
                \def\cm{{\cal M}}
                \def\cn{{\cal N}}
                \def\cf{{\cal F}}
                \def\d{\delta}
                \def\D{\Delta}
                \def\eps{\epsilon}
                \def\g{\gamma}
                \def\G{\Gamma}
                \def\vp{\varphi}
                \def\grad{\nabla}
                \def\h{ {1\over 2}  }
                \def\hc{\hat{c}}
                \def\hd{\hat{d}}
                \def\hg{\hat{g}}
                \def\/{\frac}
                \def\hp{ {+{1\over 2}}  }
                \def\hm{ {-{1\over 2}}  }
                \def\k{\kappa}
                \def\l{\lambda}
                \def\L{\Lambda}
                \def\lg{\langle}
                \def\m{\mu}
                \def\n{\nu}
                \def\o{\omega}
                \def\O{\Omega}
                \def\p{\phi}
                \def\pa{\partial}
                \def\pr{\prime}
                \def\qq{\qquad}
                \def\ra{\rightarrow}
                \def\rh{\rho}
                \def\vp{\varphi}
                \def\rg{\rangle}
                \def\s{\sigma}
                \def\t{\tau}
                \def\th{\theta}
                \def\ti{\tilde}
                \def\u{\upsilon}
                \def\wti{\widetilde}
                \def\inte{\int dx }
                \def\xb{\bar{x}}
                \def\yb{\bar{y}}
                \def\({\Big(}
                \def\){\Big)}
                \def\[{\Big[}
                \def\]{\Big]}

                \def\rlx{\relax\leavevmode}
                \def\inbar{\vrule height1.5ex width.4pt depth0pt}
                \def\IZ{\rlx\hbox{\sf Z\kern-.4em Z}}
                \def\IR{\rlx\hbox{\rm I\kern-.18em R}}
                \def\IC{\rlx\hbox{\,$\inbar\kern-.3em{\rm C}$}}
                \def\IN{\rlx\hbox{\rm I\kern-.18em N}}
                \def\IO{\rlx\hbox{\,$\inbar\kern-.3em{\rm O}$}}
                \def\IP{\rlx\hbox{\rm I\kern-.18em P}}
                \def\IQ{\rlx\hbox{\,$\inbar\kern-.3em{\rm Q}$}}
                \def\IF{\rlx\hbox{\rm I\kern-.18em F}}
                \def\IG{\rlx\hbox{\,$\inbar\kern-.3em{\rm G}$}}
                \def\IH{\rlx\hbox{\rm I\kern-.18em H}}
                \def\II{\rlx\hbox{\rm I\kern-.18em I}}
                \def\IK{\rlx\hbox{\rm I\kern-.18em K}}
                \def\IL{\rlx\hbox{\rm I\kern-.18em L}}
                \def\one{\hbox{{1}\kern-.25em\hbox{l}}}
                \def\0#1{\relax\ifmmode\mathaccent"7017{#1}%
                B        \else\accent23#1\relax\fi}
                \def\omz{\0 \omega}
                %

                \def\tr{\mathop{\rm tr}}
                \def\Tr{\mathop{\rm Tr}}
                \def\partder#1#2{{\partial #1\over\partial #2}}
                \def\ds{{\cal D}_s}
                \def\wtwo{{\wti W}_2}
\vspace{.2in}
\begin{center}
{\large\bf Heteroscedastic controlled calibration model applied to analytical chemistry}
\end{center}

\vspace{1in}

\begin{center}

Betsab\'e G. Blas Achic and M\^onica C. Sandoval

\vspace{.5 cm} \small

\par \vskip .1in \noindent
Departamento de Estat\'{\i}stica, Universidade de S\~ao Paulo, S\~ao Paulo, Brasil\\
\normalsize
\end{center}
\vspace{1 cm}

\begin{abstract}
In chemical analysis made by laboratories one has the problem
of determining the concentration of a chemical element
in a sample. In order to tackle this problem the guide EURACHEM/CITAC recommends the application of the linear calibration model, so implicitly assume that
there is no measurement error in the independent
variable $X$. In this work, it is proposed a new calibration
 model
 assuming that the independent variable is
 controlled.
This assumption is appropriate in
 chemical analysis
 where the process tempting to attain the fixed
known
 value $X$ generates an
 error and the resulting value is $x$, which is
 not an
 observable. However, observations on
its surrogate $X$ are available. A simulation study is
carried out in order to verify some properties of  the
estimators derived for the new model and it is also considered
the usual calibration model to compare it with the new
approach. Three applications are considered to verify
the performance of the new approach.

 \vskip .3in

\end{abstract}

\par \vskip .3in \noindent
\vspace{1 cm}

\section{Introduction}

The usual calibration model \cite{shukla} is commonly used to estimate the concentration $X_{0}$ of a chemical species in a test sample. It typically assumes that the independent variable is fixed and it is not subject to error. However, in applications in analytical chemistry this variable is subject to error which arises from the preparation process of a standard solution. In many studies, such as \cite{eura}, \cite{Queenie} and \cite{lutz}, it is attempted to consider the uncertainties due to the preparation process of the standard solutions by application of the error propagation law to the standard error of the estimator of $X_{0}$.

We have that the concentration of the standard solution is pre-fixed by the chemical analyst and a process is carried out attempting to attain it, this process generates errors. Hence, in this case it arises the so called controlled variable \cite{berk50}, where the controlled variable $X$ is defined by the pre-fixed concentration value of the standard solution  which is expressed by the equation $X=x+\delta$, where $x$ is the unobserved variable and $\delta$ is the measurement error variable.

In \cite{bmo} it was proposed the so called homoscedastic controlled calibration model. This model is formulated in the framework of the usual calibration model assuming that the independent variable is a controlled variable and the associated measurement errors have \textit{equal} variances.

In \cite{ballico} and \cite{bet}, some methods to compute the uncertainties in certain values obtained through measurements are studied. In \cite{bet}, the uncertainties of standard solutions are computed  and it is observed that these uncertainties depend on the concentration values, so we can observe that the usual calibration model and the homoscedastic controlled calibration model seem not to be the more suitable ones. This problem motivates us to study a calibration model that considers the errors variability of the preparation of standard solutions. In this work we propose a calibration model that incorporates the errors variability arisen from the preparation process of the standard solution and we call it as \textit{ the  heteroscedastic controlled calibration model}. This work is a continuation to  our previous paper \cite{bmo} in which it was undertaken the study of the  so-called  homoscedastic controlled calibration model which assumed \textit{equal} variance errors.

The paper is organized as follows. In Section \ref{model}, we formulate the heteroscedastic controlled calibration model. In Section \ref{simulation}, a simulation study to test the new approach is presented. In Section \ref{applications}, three applications are considered which show that the proposed model seems to be more adequate. Section \ref{remarks} presents our concluding remarks. Finally, we present in Appendix \ref{appa} the usual calibration model, and in Appendix \ref{appb} some tables showing the results of the simulation study.\\

\section{The proposed model}
\label{model}

Among the relevant problems in chemical analysis is the one related to the estimation of the concentration $X_0$ of a chemical compound in a given sample. In order to tackle this problem it is used a statistical calibration model, which is defined by a two-step process. This problem has been considered in \cite{tallis} and \cite{lwin}.

The first stage of the calibration model is given by data points $(X,Y)$ which is determined in an experiment where the independent variable $X$ is the one that the experimenter selects. For instance, the concentrations of the standard solutions that a chemist prepares are independent variables since any concentration may be chosen. The dependent variable $Y$ is a measurable property of the independent variable. For example, the dependent variable may be the amount of intensity supplied by the plasma spectrometry method, since the intensity  depends on the concentration.

In the second stage of the calibration model it is prepared a suitable sample related to the unknown concentration $X_0$ in order to obtain the measurements $Y_0$.

We have that the standard concentration $X$ is  fixed by the analyst and the process of preparation attempting to get it produces an error $\d$, and the unobserved quantity attained is $x$. Considering the usual calibration model defined by the equations \ref{m1} and \ref{m2} in the Appendix \ref{appa} and the equation $X=x+\d$, we define the heteroscedastic controlled calibration model as
\begin{eqnarray}
\label{m11}
Y_{i}&=&\alpha+\beta x_{i}+\epsilon_{i},\,\,\,\,\,i=1,2\cdots,n,\\
\label{22}
X_i&=&x_i+\d_i,\,\,\,\,\,\,\,\,\,\,\,\,\,\,\,\,\,\,\,\,i=1,2\cdots,n,\\
\label{m33}
Y_{0i}&=&\alpha+\beta X_{0}+\epsilon_{i},\,\,\,\,i=n+1,n+2,\cdots,n+k.
\end{eqnarray}
It is considered the usual calibration model assumptions (see Appendix A) in addition to the following conditions
\begin{itemize}
   \item $\delta_{1},\delta_{2},\cdots,\delta_{n}$ are independent and normally distributed with mean 0.
   \item the variances $\sigma_{\delta_{i}}^{2},\,(i=1,\cdots,n)$ are supposed to be known.
  \item $\delta_{i},\,i=1,\cdots,n$ and $\epsilon_{i},\,i=1,\cdots,n+k$ are independent.
\end{itemize}
Observe that in the model described above we only consider the case when the variances $\sigma_{\delta_{i}}^{2},\,\,\,i=1,\cdots,n$ are known. It is a generalization of the homoscedastic controlled calibration model discussed in \cite{bmo}, when it is considered
$\sigma^{2}_{\delta_{i}}=\sigma_{\delta}^{2}$ for all $i$ and  the known $\sigma_{\delta}^{2}$ case. This new model is also a generalization of the usual calibration model in which one takes $\delta_{i}=0,\,i=1,\cdots,n$.

For the heteroscedastic controlled calibration model the logarithm of the likelihood function is given by
{\small\begin{eqnarray}
\nonumber
l(\,\alpha,\beta,X_{0},\sigma_{\epsilon}^{2})&&\propto-\frac{1}{2}\sum_{i=1}^{n}log(\gamma_i)-\frac{k}{2}log(\sigma_{\epsilon}^{2})\\
\label{L1}
&&-\frac{1}{2}\left[\sum_{i=1}^{n}\frac{(Y_{i}-\alpha-\beta
X_{i})^{2}}{\gamma_i}+\sum_{i=n+1}^{n+k}\frac{(Y_{0i}-\alpha-\beta
X_{0})^{2}}{\sigma_{\epsilon}^{2}}\right],
\end{eqnarray}}
where $\gamma_i=\sigma_{\epsilon}^{2}+\beta^{2}\sigma_{\delta_{i}}^{2}$, $i=1,\cdots,n$.
Solving $\partial l /\partial\alpha=0$ and $\partial l/\partial X_{0}=0$ one can get the maximum likelihood estimator of $\alpha$ and $X_{0}$ given,
respectively, by
\begin{equation}
\label{axo}
\hat{\alpha}=\bar{Y}-\hat{\beta}\bar{X}\,\,\,\,\,\mbox{and}\,\,\,\,\,\hat{X}_{0}=\frac{\bar{Y}_{0}-\hat{\alpha}}{\hat{\beta}}.
\end{equation}
From (\ref{L1}) and (\ref{axo}), it follows that the logarithm of the likelihood function for $(\alpha,\beta,X_0,\s_{\eps}^2)$ can be writen as
{\small\begin{eqnarray}
\label{L2}
l(\alpha,\beta,X_{0},\sigma_{\epsilon}^{2})&&\propto-\frac{1}{2}\sum_{i=1}^{n}log(\gamma_i)-\frac{k}{2}log(\sigma_{\epsilon}^{2})\\
\nonumber
&&-\frac{1}{2}\left[\sum_{i=1}^{n}\frac{[(Y_{i}-\bar{Y})-\beta
(X_{i}-\bar{X})]^{2}}{\gamma_i}+\frac{1}{\sigma_{\epsilon}^{2}}\sum_{i=n+1}^{n+k}(Y_{0i}-\bar{Y_{0}})^{2}\right].
\end{eqnarray}}
Making $\partial l /\partial\beta=0$, $\partial l/\partial\sigma_{\epsilon}^{2}=0$ in the logarithm of the likelihood function (\ref{L2}), we have the following equations
\begin{eqnarray}
\label{ebh}
\sum_{i=1}^{n}\frac{\beta\sigma_{\delta_{i}}^{2}\left[\gamma_i-(Y_{i}-\alpha-\beta
 X_{i})^{2}\right]}{\gamma_i^{2}}=
 \sum_{i=1}^{n}\frac{X_{i}(Y_{i}-\alpha-\beta
 X_{i})}{\gamma_i}\\
\label{eeh}
\sum_{i=1}^{n}\frac{\gamma_i-(Y_{i}-\alpha-\beta
 X_{i})^{2}}{\gamma_i^{2}}=\sum_{i=n+1}^{n+k}\frac{(Y_{0i}-\bar{Y_{0}})^{2}}{\sigma_{\epsilon}^{4}}-\frac{k}{\sigma_{\epsilon}^{2}}.
\end{eqnarray}
The estimates of $\beta$ and $\sigma_{\epsilon}^{2}$ can be
obtained through some iterative method that solves the equations (\ref{ebh}) and (\ref{eeh}).

The  Fisher expected information $I(\theta)=I(\alpha,\beta,X_{0},\sigma_{\epsilon}^{2})$ is given by\\

$I(\theta)= \left(\begin{array}{ccccc}
    \sum_{i=1}^{n} \frac{1}{\gamma_{i}}+\frac{k}{\sigma_{\epsilon}^{2}} &\sum_{i=1}^{n} \frac{X_{i}}{\gamma_{i}}+ \frac{kX_{0}}{\sigma_{\epsilon}^{2}} &\frac{k\beta}{\sigma_{\epsilon}^{2}}  &0 \\
    \sum_{i=1}^{n} \frac{X_{i}}{\gamma_{i}}+ \frac{kX_{0}}{\sigma_{\epsilon}^{2}}  &  \sum_{i=1}^{n}\frac{X_{i}^{2}}{\gamma_{i}}+2\beta^{2}\sum_{i=1}^{n}\frac{\sigma_{\delta_{i}}^{4}}{\gamma_{i}^{2}}+\frac{kX_{0}^{2}}{\sigma_{\epsilon}^{2}}  & \frac{k\beta X_{0}}{\sigma_{\epsilon}^{2}}  & \beta\sum_{i=1}^{n}\frac{\sigma_{\delta_{i}}^{2}}{\gamma_{i}^{2}}\\
      \frac{k\beta}{\sigma_{\epsilon}^{2}}&  \frac{k\beta X_{0}}{\sigma_{\epsilon}^{2}}  &\frac{k\beta^{2}}{\sigma_{\epsilon}^{2}} &0 \\
      0 & \beta\sum_{i=1}^{n}\frac{\sigma_{\delta_{i}}^{2}}{\gamma_{i}^{2}} & 0& \sum_{i=1}^{n}\frac{1}{2\gamma_{i}^{2}}+\frac{k}{2\sigma_{\epsilon}^{4}}\\
\end{array}\right)$\\

When $k=qn$, $q\in Q^{+}$ and $n\rightarrow \infty$, the estimator $\hat{\theta}$ is approximately normally distributed with mean $\theta$ and variance $I(\theta)^{-1}$, thus the approximate variance to order $n^{-1}$ for $\hat{X}_{0}$ is given by
\begin{equation}
\label{varhete}
V(\hat{X}_{0})=\frac{\sigma_{\epsilon}^{2}}{\beta^{2}}\left[\frac{1}{n}+\frac{1}{k}-\frac{E_{1}}{n\sigma_{\epsilon}^{2}E_{2}}\right],
\end{equation}
where
\begin{eqnarray}
\nonumber
E_{1}&=&-n\sum_{i=1}^{n}\frac{X_{0}^{2}\sigma_{\epsilon}^{4}}{\gamma_{i}}\sum_{i=1}^{n}\frac{1}{\gamma_{i}^{2}}-nk\sum_{i=1}^{n}\frac{X_{0}^{2}}{\gamma_{i}}-n\sum_{i=1}^{n}\frac{X_{i}^{2}\sigma_{\epsilon}^{4}}{\gamma_{i}}\sum_{i=1}^{n}\frac{1}{\gamma_{i}^{2}}-nk\sum_{i=1}^{n}\frac{X_{i}^{2}}{\gamma_{i}}\\
\nonumber
&&-2n\beta^{2}\sum_{i=1}^{n}\frac{\sigma_{\delta_{i}}^{4}\sigma_{\epsilon}^{4}}{\gamma_{i}^{2}}\sum_{i=1}^{n}\frac{1}{\gamma_{i}^{2}}-2nk\beta^{2}\sum_{i=1}^{n}\frac{\sigma_{\delta_{i}}^{4}}{\gamma_{i}^{2}}+2n\beta^{2}\sigma_{\epsilon}^{4}\left[\sum_{i=1}^{n}\frac{\sigma_{\delta_{i}}^{2}}{\gamma_{i}^{2}}\right]^{2}\\
\nonumber
&&+2nX_{0}\sigma_{\epsilon}^{4}\sum_{i=1}^{n}\frac{X_{i}}{\gamma_{i}}\sum_{i=1}^{n}\frac{1}{\gamma_{i}^{2}}+2nkX_{0}\sum_{i=1}^{n}\frac{X_{i}}{\gamma_{i}}+\sigma_{\epsilon}^{6}\sum_{i=1}^{n}\frac{X_{i}^{2}}{\gamma_{i}}\sum_{i=1}^{n}\frac{1}{\gamma_{i}^{2}}\sum_{i=1}^{n}\frac{1}{\gamma_{i}}\\
\nonumber
&&+k\sigma_{\epsilon}^{2}\sum_{i=1}^{n}\frac{X_{i}^{2}}{\gamma_{i}}\sum_{i=1}^{n}\frac{1}{\gamma_{i}}+2\beta^{2}\sigma_{\epsilon}^{6}\sum_{i=1}^{n}\frac{\sigma_{\delta_{i}}^{4}}{\gamma_{i}^{2}}\sum_{i=1}^{n}\frac{1}{\gamma_{i}^{2}}\sum_{i=1}^{n}\frac{1}{\gamma_{i}}+2k\beta^{2}\sigma_{\epsilon}^{2}\sum_{i=1}^{n}\frac{1}{\gamma_{i}}\sum_{i=1}^{n}\frac{\sigma_{\delta_{i}}^{4}}{\gamma_{i}^{2}}\\
\nonumber
&&-2\beta^{2}\sigma_{\epsilon}^{6}\left[\sum_{i=1}^{n}\frac{\sigma_{\delta_{i}}^{2}}{\gamma_{i}^{2}}\right]^{2}\sum_{i=1}^{n}\frac{1}{\gamma_{i}}-\sigma_{\epsilon}^{6}\left[\sum_{i=1}^{n}\frac{X_{i}}{\gamma_{i}}\right]^{2}\sum_{i=1}^{n}\frac{1}{\gamma_{i}^{2}}-k\sigma_{\epsilon}^{2}\left[\sum_{i=1}^{n}\frac{X_{i}}{\gamma_{i}}\right]^{2}
\end{eqnarray}
and
\begin{eqnarray}
\nonumber
E_{2}&=&\sigma_{\epsilon}^{4}\sum_{i=1}^{n}\frac{X_{i}^{2}}{\gamma_{i}}\sum_{i=1}^{n}\frac{1}{\gamma_{i}^{2}}\sum_{i=1}^{n}\frac{1}{\gamma_{i}}+k\sum_{i=1}^{n}\frac{X_{i}^{2}}{\gamma_{i}}\sum_{i=1}^{n}\frac{1}{\gamma_{i}}+2\sigma_{\epsilon}^{4}\beta^{2}\sum_{i=1}^{n}\frac{\sigma_{\delta_{i}}^{4}}{\gamma_{i}^{2}}\sum_{i=1}^{n}\frac{1}{\gamma_{i}^{2}}\sum_{i=1}^{n}\frac{1}{\gamma_{i}}\\
\nonumber
&&+2k\beta^{2}\sum_{i=1}^{n}\frac{\sigma_{\delta_{i}}^{4}}{\gamma_{i}^{2}}\sum_{i=1}^{n}\frac{1}{\gamma_{i}}-2\beta^{2}\sigma_{\epsilon}^{4}\left[\sum_{i=1}^{n}\frac{\sigma_{\delta_{i}}^{2}}{\gamma_{i}^{2}}\right]^{2}\sum_{i=1}^{n}\frac{1}{\gamma_{i}}\\
\nonumber
&&-\sigma_{\epsilon}^{4}\left[\sum_{i=1}^{n}\frac{X_{i}}{\gamma_{i}}\right]^{2}\sum_{i=1}^{n}\frac{1}{\gamma_{i}^{2}}-k\left[\sum_{i=1}^{n}\frac{X_{i}}{\gamma_{i}}\right]^{2}.
\end{eqnarray}

Note that when $\sigma_{\delta_{i}}^{2}=0,\,i=1,\cdots,n,$  the
expression (\ref{varhete}) is reduced to the variance of the usual model given in (\ref{veurachem}) and when
$\sigma_{\delta_{i}}^{2}=\sigma_{\delta}^{2}$ (for all $i$) the
expression (\ref{varhete}) is also reduced to the variance of the homoscedastic model
when $\sigma_{\delta}^{2}$ is known (see eq. (2.12)  of ref. \cite{bmo}).

In order to construct a confidence interval for $X_0$ we consider the interval (\ref{uinterv}), where $\hat{V}(\hat{X}_{0C})$ is the estimated variance that follows from (\ref{varhete}).

\section{Simulation study}
\label{simulation}

We present a simulation study to compare the performance of the estimators obtained from the heteroscedastic
controlled calibration model (Proposed-M) with the results obtained by  considering the usual model (Usual-M).

It was considered 3000 samples generated from the Proposed-M. In all the  samples, the parameters $\alpha$ and $\beta$
 take the values 0.1 and 2, respectively. The range of values for the controlled variable was
[0,2]. The fixed values for the controlled variable were
$x_{1}=0,\,x_{i}=x_{i-1}+2/(n-1),\,i=2,\cdots,n,$ and the parameter values
$X_{0}$ were 0.01 (extreme inferior value), 0.8 (near to the central value) and 1.9
(extreme superior value).  For the first and second stages we consider the sample of sizes $n=5,\,20,\,100,\,5000$ and $k=2,\,20,\,100,\,500$, respectively. We consider $\sigma_{\epsilon}^{2}=0.04$ and the
maximum parameter values of $\sigma_{\delta}^{2}$ as $max\{\sigma_{\delta_i}^{2}\}_{i=1}^n$= 0.1. We consider  $\sigma_{\delta_i}^{2}=i\times 0.1/n\,\,\,\mbox{for } i=1,\cdots,n$.

The empirical mean bias is given by $\sum_{j=1}^{3000}(\hat{X}_{0}-X_{0})/3000$  and the empirical
mean squared error (MSE) is given by $\sum_{j=1}^{3000}(\hat{X}_{0}-X_{0})^{2}/3000$. The mean estimated variance
 of $\hat{X}_{0}$ is given by $\sum_{j=1}^{3000}\hat{V}(\hat{X}_{0})/3000$. The theoretical variances of $\hat{X}_{0}$ is  referred to the expressions (\ref{veurachem}) and (\ref{varhete}) evaluated on the relevant parameter values. In Appendix B it is presented the simulation results.

In Table \ref{pd0p}, we observe that, in general, the bias of $\hat{X}_{0}$ from the usual model is smaller than the value supplied by the proposed model, but related to the MSE we have that the outcome from usual model is greater compared with MSE of the proposed model. Also, we observe that the mean estimated variance from the proposed model is closer to the theoretical variance as compared to the outcome from the usual model.

Analyzing Table \ref{pd1p}, we observe that the amplitude of the proposed model,
in most cases, is smaller when compared with the estimate of the usual model. For all $n$ and $X_0$ the amplitude from the usual model greatly decreases as the size of k increases,  this behavior is being reflected on the covering percentage decreasing to less than 95\%. Adopting the correct model we have that when $k$ increases the confidence interval amplitude decreases and the covering percentage increases approaching 95\%.

\label{appb}
\begin{table}[!th]
\caption{Empirical bias and mean squared error, the mean
estimated variance and theoretical variance of $\hat{X}_{0}$.}\label{pd0p}
\begin{center}
\begin{scriptsize}
\begin{tabular}{r|r|r|cc|cc|c|c|c}\hline
$X_0$ & n & k &\multicolumn{2}{c|}{Usual-M} &\multicolumn{2}{c|}{Proposed-M}&Usual-M&Proposed-M& Theorical variance\\\cline{4-10}
 &&& Bias & MSE & Bias & MSE &$\hat{V}(\hat{X}_{0})$ & $\hat{V}(\hat{X}_{0})$ & $V(\hat{X}_{0})$ \\\hline
 0.01   &   5   &   2   &   -0.0156 &   0.0350  &   -0.0318 &   0.0334  &   0.0398  &   0.0143  &   0.0257  \\
    &       &   20  &   -0.0236 &   0.0319  &   -0.0445 &   0.0278  &   0.0131  &   0.0156  &   0.0211  \\
    &       &   100 &   -0.0183 &   0.0306  &   -0.0429 &   0.0276  &   0.0084  &   0.0155  &   0.0207  \\
    &   20  &   2   &   -0.0076 &   0.0119  &   -0.0049 &   0.0100  &   0.0365  &   0.0053  &   0.0097  \\
    &       &   20  &   -0.0055 &   0.0074  &   -0.0073 &   0.0055  &   0.0081  &   0.0036  &   0.0051  \\
    &       &   100 &   -0.0059 &   0.0068  &   -0.0101 &   0.0050  &   0.0036  &   0.0033  &   0.0047  \\
    &   100 &   2   &   0.0003  &   0.0063  &   0.0020  &   0.0059  &   0.0315  &   0.0047  &   0.0059  \\
    &       &   20  &   -0.0023 &   0.0019  &   -0.0020 &   0.0014  &   0.0046  &   0.0011  &   0.0014  \\
    &       &   100 &   -0.0014 &   0.0015  &   -0.0013 &   0.0010  &   0.0017  &   0.0007  &   0.0010  \\
    &   5000    &   2   &   0.0008  &   0.0055  &   0.0008  &   0.0055  &   0.0300  &   0.0050  &   0.0050  \\
    &       &   20  &   0.0000  &   0.0005  &   0.0000  &   0.0005  &   0.0030  &   0.0005  &   0.0005  \\
    &       &   100 &   -0.0003 &   0.0001  &   -0.0004 &   0.0001  &   0.0006  &   0.0001  &   0.0001  \\
    &       &   500 &   0.0000  &   0.0000  &   0.0000  &   0.0000  &   0.0002  &   0.0000  &   0.0000  \\  \hline
0.8 &   5   &   2   &   0.0061  &   0.0193  &   0.0025  &   0.0202  &   0.0254  &   0.0089  &   0.0167  \\
    &       &   20  &   0.0033  &   0.0135  &   -0.0019 &   0.0139  &   0.0047  &   0.0081  &   0.0122  \\
    &       &   100 &   0.0014  &   0.0132  &   -0.0037 &   0.0137  &   0.0029  &   0.0078  &   0.0118  \\
    &   20  &   2   &   0.0015  &   0.0077  &   0.0015  &   0.0077  &   0.0291  &   0.0042  &   0.0074  \\
    &       &   20  &   0.0016  &   0.0032  &   0.0009  &   0.0032  &   0.0036  &   0.0020  &   0.0029  \\
    &       &   100 &   -0.0005 &   0.0026  &   -0.0018 &   0.0026  &   0.0012  &   0.0016  &   0.0025  \\
    &   100 &   2   &   0.0010  &   0.0055  &   0.0014  &   0.0054  &   0.0299  &   0.0044  &   0.0055  \\
    &       &   20  &   0.0006  &   0.0010  &   0.0007  &   0.0010  &   0.0031  &   0.0008  &   0.0010  \\
    &       &   100 &   -0.0001 &   0.0006  &   -0.0001 &   0.0006  &   0.0007  &   0.0004  &   0.0006  \\
    &   5000    &   2   &   0.0014  &   0.0051  &   0.0014  &   0.0051  &   0.0300  &   0.0050  &   0.0050  \\
    &       &   20  &   0.0006  &   0.0005  &   0.0006  &   0.0005  &   0.0030  &   0.0005  &   0.0005  \\
    &       &   100 &   0.0001  &   0.0001  &   0.0001  &   0.0001  &   0.0006  &   0.0001  &   0.0001  \\
    &       &   500 &   0.0000  &   0.0000  &   0.0000  &   0.0000  &   0.0001  &   0.0000  &   0.0000  \\  \hline
1.9 &   5   &   2   &   0.0500  &   0.0802  &   0.0582  &   0.0704  &   0.0432  &   0.0275  &   0.0482  \\
    &       &   20  &   0.0314  &   0.0620  &   0.0503  &   0.0562  &   0.0124  &   0.0278  &   0.0435  \\
    &       &   100 &   0.0434  &   0.0645  &   0.0594  &   0.0587  &   0.0085  &   0.0283  &   0.0430  \\
    &   20  &   2   &   0.0070  &   0.0213  &   0.0054  &   0.0185  &   0.0351  &   0.0086  &   0.0166  \\
    &       &   20  &   0.0117  &   0.0160  &   0.0127  &   0.0132  &   0.0075  &   0.0067  &   0.0118  \\
    &       &   100 &   0.0099  &   0.0161  &   0.0104  &   0.0130  &   0.0033  &   0.0064  &   0.0114  \\
    &   100 &   2   &   0.0016  &   0.0080  &   0.0007  &   0.0076  &   0.0312  &   0.0055  &   0.0074  \\
    &       &   20  &   0.0019  &   0.0035  &   0.0014  &   0.0029  &   0.0043  &   0.0017  &   0.0028  \\
    &       &   100 &   0.0001  &   0.0031  &   0.0009  &   0.0025  &   0.0015  &   0.0013  &   0.0024  \\
    &   5000    &   2   &   -0.0008 &   0.0051  &   -0.0009 &   0.0051  &   0.0300  &   0.0050  &   0.0050  \\
    &       &   20  &   -0.0003 &   0.0006  &   -0.0004 &   0.0006  &   0.0030  &   0.0005  &   0.0005  \\
    &       &   100 &   -0.0003 &   0.0002  &   -0.0002 &   0.0001  &   0.0006  &   0.0001  &   0.0001  \\
    &       &   500 &   0.0000  &   0.0001  &   0.0000  &   0.0001  &   0.0001  &   0.0000  &   0.0001  \\
  \hline
\end{tabular}
\end{scriptsize}
\end{center}
\end{table}

\newpage
\begin{table}[!t]
\caption{Covering percentage (\%) and amplitude (A) of the intervals with a 95\%
confidence level for the parameter $X_{0}$.} \label{pd1p}
\begin{scriptsize}
\begin{center}
\begin{tabular}{r|r|r|cc|cc}\hline
$X_{0}$ & $n$&$k$&\multicolumn{2}{c|}{Usual-M}&\multicolumn{2}{c}{Proposed-M}\\\cline{4-7}
 &&& \%&$A$&\%&$A$\\
\hline
0.01    &   5   &   2   &   89  &   0.35    &   78  &   0.22    \\
    &       &   20  &   78  &   0.21    &   89  &   0.24    \\
    &       &   100 &   70  &   0.17    &   88  &   0.24    \\
    &   20  &   2   &   100 &   0.37    &   74  &   0.13    \\
    &       &   20  &   95  &   0.17    &   90  &   0.12    \\
    &       &   100 &   85  &   0.12    &   90  &   0.11    \\
    &   100 &   2   &   100 &   0.35    &   84  &   0.13    \\
    &       &   20  &   100 &   0.13    &   91  &   0.06    \\
    &       &   100 &   96  &   0.08    &   90  &   0.05    \\
    &   5000    &   2   &   100 &   0.34    &   94  &   0.14    \\
    &       &   20  &   100 &   0.11    &   95  &   0.04    \\
    &       &   100 &   100 &   0.05    &   94  &   0.02    \\
    &       &   500 &   100 &   0.02    &   94  &   0.01    \\\hline
0.8 &   5   &   2   &   90  &   0.28    &   78  &   0.18    \\
    &       &   20  &   73  &   0.13    &   87  &   0.17    \\
    &       &   100 &   63  &   0.10    &   87  &   0.17    \\
    &   20  &   2   &   100 &   0.33    &   73  &   0.11    \\
    &       &   20  &   95  &   0.12    &   87  &   0.09    \\
    &       &   100 &   81  &   0.07    &   88  &   0.08    \\
    &   100 &   2   &   100 &   0.34    &   86  &   0.12    \\
    &       &   20  &   100 &   0.11    &   91  &   0.05    \\
    &       &   100 &   97  &   0.05    &   89  &   0.04    \\
    &   5000    &   2   &   100 &   0.34    &   95  &   0.14    \\
    &       &   20  &   100 &   0.11    &   95  &   0.04    \\
    &       &   100 &   100 &   0.05    &   95  &   0.02    \\
    &       &   500 &   100 &   0.02    &   93  &   0.01    \\\hline
1.9 &   5   &   2   &   78  &   0.35    &   81  &   0.31    \\
    &       &   20  &   59  &   0.20    &   86  &   0.32    \\
    &       &   100 &   51  &   0.17    &   87  &   0.32    \\
    &   20  &   2   &   98  &   0.36    &   78  &   0.17    \\
    &       &   20  &   81  &   0.17    &   84  &   0.16    \\
    &       &   100 &   62  &   0.11    &   84  &   0.16    \\
    &   100 &   2   &   100 &   0.34    &   87  &   0.14    \\
    &       &   20  &   97  &   0.13    &   87  &   0.08    \\
    &       &   100 &   83  &   0.08    &   84  &   0.07    \\
    &   5000    &   2   &   100 &   0.34    &   95  &   0.14    \\
    &       &   20  &   100 &   0.11    &   94  &   0.04    \\
    &       &   100 &   100 &   0.05    &   93  &   0.02    \\
    &       &   500 &   99  &   0.02    &   89  &   0.01    \\
\hline
\end{tabular}
\end{center}
\end{scriptsize}
\end{table}

\section{Application}
\label{applications}

In this section we illustrate the usefulness of the proposed model by applying it to the data supplied by the chemical laboratory of the ``Instituto de Pesquisas Tecnol\'ogicas do  Estado de S\~ao Paulo (IPT)" - Brasil. The outcome from the proposed approach are also compared with the results from the usual model. Our main interest is
to estimate the unknown concentration value $X_{0}$ of a sample of the
chemical elements such as chromium, cadmium and lead.

Table \ref{cr0} below presents the fixed values of concentration for the
standard solutions with their related uncertainty ($u(X_{i})$) and the corresponding intensities for the chromium, cadmium and lead elements. The uncertainties considered are computed using the method recommended by the ISOGUM guide (see \cite{isogum}) and the intensities are supplied by the plasma spectrometry method. This data is referred to the first stage of the
heteroscedastic controlled calibration model.

Moreover, Table \ref{cr01} below presents the intensities of the sample solutions of chromium, cadmium and lead elements. These data are referred to as the second stage of the calibration model.

Observing Tables \ref{cr0} and \ref{cr01} we verify that the uncertainty  values increase with the concentration values.

We consider $\sigma_{\delta_{i}}^{2}=u(X_{i})^{2}$. The expanded uncertainty $U(X_{0})$ is obtained multiplying the squared root of the estimate of variance of $\hat{X}_{0}$ by the value 1.96 (see \cite{eura} and \cite{bet}).

We use the \textsl{optim} command from R-project program to estimate the parameters $\beta$ and $\sigma_{\epsilon}^{2}$ on the likelihood function of the proposed model (\ref{L2}). We use as initial point the estimates from $\hat{\beta}=\sum_{i=1}^{n}(X_i-\bar{x})(Y_i-\bar{Y})/\sum_{i=1}^{n}(X_i-\bar{X})^2$ and $\hat{\sigma}_{\epsilon}^{2}=\sum_{i=1}^{n}(Y_{0i}-\bar{Y_0})^2/n$, which are the estimators from the homoscedastic controlled calibration model when $\sigma_{\delta}^{2}$ is unknow \cite{bmo}.

Table \ref{cr1} presents estimates of $\alpha,\,\beta,\,X_{0},$ $V(\hat{X}_{0})$ and the expanded uncertainty, $U(X_{0})$, from the proposed model (Proposed-M) of chromium, cadmium and lead elements. Also, we present the estimates obtained from usual calibration model (Usual-M) to observe the performance of both models.

In Table \ref{cr1}, for cadmium and lead elements, we observe that the estimates of $\alpha$, $\beta$ and $X_0$ from the Proposed-M and Usual-M are the same. For the chromium element, there are small differences. Also, we observe that for the chromium element there is a small difference between the estimates of $X_{0}$ and $U(X_{0})$ respectively obtained from the usual model and the proposed model. Despite the relevant estimates of $\alpha,\beta$ and $X_0$ from both approaches for cadmium and lead element match, the  estimates of $V(X_{0})$ and $U(X_{0})$ differ considerably, the estimates obtained adopting the usual model is greater than the estimates outcome supplied by the proposed model.
\begin{table}[!t]
\caption{Concentration $(mg/g)$, uncertainty($u(X_{i})$) and intensity of the standard solutions of chromium, cadmium  and lead elements.}
\label{cr0}
\begin{center}
\begin{scriptsize}
\begin{tabular}{|c|r|r|c|r|r|c|r|r|}\hline
\multicolumn{3}{|c|}{Chromium element}&\multicolumn{3}{|c|}{Cadmium element}&\multicolumn{3}{|c|}{lead element}\\\hline
$X_{i}$&$u(X_{i})$&Intensity &$X_{i}$&$u(X_{i})$&Intensity& $X_{i}$&$u(X_{i})$&Intensity \\\hline
0.05   &0.00016&6455.900    &   0.05&0.00016    &4.89733        & 0.05&  0.00015 & 0.9471\\
0.11   &0.00027&13042.933 & 0.10&0.00027     &9.706         &0.10& 0.00025& 1.46833\\
0.26   &0.00040&32621.733 & 0.25&0.00041     &23.41333  &0.26& 0.00039& 3.09033\\
0.79   &0.00122&97364.500 & 0.73&0.00122    &69.73          &0.77& 0.00117& 8.40533\\
1.05   &0.00161&129178.100& 1.01&0.00168     &96.85667  &1.01& 0.00155&10.92667\\\hline
\end{tabular}
\end{scriptsize}
\end{center}
\end{table}

\vspace{0.01cm}
\begin{center}
\begin{table}[!t]
\begin{scriptsize}
\caption{Intensity of the sample solutions of chromium, cadmium and lead elements.} \label{cr01}
\begin{center}
\begin{tabular}{|c|c|c|}\hline
Chromium element &Cadmium element& Lead element\\\hline
   10173.6& 5.066 &1.303  \\
   10516.9& 5.027  &1.290\\
    10352.2&   5.085  &1.341\\\hline
\end{tabular}
\end{center}
\end{scriptsize}
\end{table}
\end{center}

\begin{table}[!t]
\begin{center}
\caption{Estimates of $\alpha,\,\beta,\,X_{0},$ $V(\hat{X}_{0})$ and $U(X_{0})$ related to usual and heteroscedastic model, for the chromium, cadmium and lead element.}
\label{cr1}
\begin{scriptsize}
\begin{tabular}{|c|c|c|c|c|c|c|}\hline
&\multicolumn{2}{c|}{Chromium element}&\multicolumn{2}{c|}{Cadmium element}&\multicolumn{2}{c|}{Lead element}\\\cline{2-7}
 Parameters&Usual-M&Proposed-M& Usual-M
 &Proposed-M &Usual-M&Proposed-M\\\hline
$\alpha$ & 134.9469&    124.2801&   0.454801&   0.454801&   -0.3822126& -0.3822126\\
$\beta$ &123003.7&  123027.3&   10.54381&   10.54381&   94.29881&   94.29881\\
$X_{0}$&0.08302691& 0.08309769& 0.08123556& 0.08123556&     0.05770535& 0.05770535\\
$V(\hat{X}_{0})$& 4.357870e-06&     4.474395e-06&   7.898643e-05&       4.440342e-06&   0.0001181068&   7.237226e-08\\
$U(X_{0})$&0.004091601& 0.004145942&    0.01741936& 0.004130135&    0.02130068& 0.000527281 \\\hline
\end{tabular}
\end{scriptsize}
\end{center}
\end{table}

\section{Concluding remarks}
\label{remarks}
The expanded uncertainty of $X_0$ from the proposed model arises from the errors appearing in the both process, the reading of equipment and the heteroscedastic error in the preparation of standard solutions. We observe that, despite the classical model only considers the error originated from equipment reading,  there are some applications in which the expanded uncertainty is greater than the one obtained through the new approach.\\

Various aspects of the model studied above deserve attention in
future research, e.g. it is not considerated the error arisen from the test sample solution preparation, the proposed model  can be studied by considering other type of distribution of the errors, such as skew normal distribution \cite{azzalini}. In
particular, one of the drawbacks of the usual model is that it does not consider the error in the independent variable, we believe that despite that this error being very small, it must be considered as an important property of the calibration model. We will concentrate on one of the problems described above in a future work. \\

\noindent {\bf Acknowledgments}

The authors are grateful to Prof. Dr. Heleno Bolfarine for carefully reading the manuscript and
Dr. Olga Satomi from "Instituto de Pesquisas Tecnológicas" - IPT.
Betsabé G. B. Achic has been supported by a grant from CNPq.

\newpage

\appendix

\section{Usual calibration model}
\label{appa}

The first and second stage equations of the usual linear calibration model are given, respectively, by
\begin{eqnarray}
\label{m1}
Y_{i}&=&\alpha+\beta x_{i}+\epsilon_{i},\,\,\,\,\,i=1,2\cdots,n,\\
\label{m2}
Y_{0i}&=&\alpha+\beta X_{0}+\epsilon_{i},\,\,\,\,\,i=n+1,n+2,\cdots,n+k.
\end{eqnarray}
It is considered the following assumptions:
\begin{itemize}
  \item $x_{1},x_{2},\cdots,x_{n}$ take fixed values, which are considered as true values.
  \item $\epsilon_{1},\epsilon_{2},\cdots,\epsilon_{n+k}$ are independent and normally distributed with mean 0 and variance $\sigma_{\epsilon}^{2}$.
\end{itemize}
The model parameters are $\alpha,\beta,X_{0}$ and $\sigma_{\epsilon}^{2}$ and the
main interest is to estimate the quantity $X_{0}$.

The maximum likelihood estimators of the usual calibration model are given by
\begin{eqnarray}
\label{xo}
\hat{\alpha}&=&\bar{Y}-\hat{\beta}\bar{x},\,\,\,\,\,\,\,\,\,\,\,\,\,\,\,\,\hat{\beta}=\frac{S_{xY}}{S_{xx}},\,\,\,\,\,\,\,\,\,\,\,\,\,\,\,\hat{X}_{0}=\frac{\bar{Y}_{0}-\hat{\alpha}}{\hat{\beta}},\\
\label{se}
\sigma_{\epsilon}^{2}&=&\frac{1}{n+k}\big[\sum_{i=1}^{n}(Y_{i}-\hat{\alpha}-\hat{\beta}x_{i})^{2}+\sum_{i=n+1}^{n+k}(Y_{0i}-\bar{Y}_{0})^{2}\big],
\end{eqnarray}
where
\begin{eqnarray}
\nonumber \bar{x}&=&\frac{1}{n}\sum_{i=1}^{n}x_{i},\,\,\,\,
\bar{Y}=\frac{1}{n}\sum_{i=1}^{n}Y_{i},\,\,\,\,
S_{xY}=\frac{1}{n}\sum_{i=1}^{n}(x_{i}-\bar{x})(Y_{i}-\bar{Y}),\\\nonumber
S_{xx}&=&\frac{1}{n}\sum_{i=1}^{n}(x_{i}-\bar{x})^{2},\,\,\,\,
\bar{Y}_{0}=\frac{1}{n}\sum_{i=n+1}^{n+k}Y_{0i}.
\end{eqnarray}
The approximation of order $n^{-1}$ for the variance of $\hat{X}_{0}$ is given by
\begin{equation}
 \label{veurachem}
V_{1}(\hat{X}_{0})=\frac{\sigma_{\epsilon}^{2}}{\beta^{2}}\left[\frac{1}{k}+\frac{1}{n}+\frac{(\bar{x}-X_{0})^{2}}{nS_{xx}}\right].
\end{equation}
In order to construct a confidence interval for $X_{0}$, we consider that
\begin{equation}
\label{pibotep}
\frac{\hat{X}_{0}-X_{0}}{\sqrt{\hat{V}(\hat{X}_{0})}}\stackrel{D}{\longrightarrow
}N(0,1),
\end{equation}
hence, the approximated
confidence interval for $X_{0}$ with a confidence level $(1-\alpha)$, is given by
\begin{equation}
\label{uinterv}
\left(\hat{X}_{0}-z_{\frac{\alpha}{2}}\sqrt{\hat{V}(\hat{X}_{0})},\hat{X}_{0}+z_{\frac{\alpha}{2}}\sqrt{\hat{V}(\hat{X}_{0})}\right),
\end{equation}
where $z_{\frac{\alpha}{2}}$ is the quantile of order
$(1-\frac{\alpha}{2})$ of the standard normal distribution.\\

\end{document}